# Submolecular-resolution non-invasive imaging of interfacial water with atomic force microscopy

Jinbo Peng[1*], Jing Guo[1*], Prokop Hapala[2*], Duanyun Cao[1], Runze Ma[1], Bowei Cheng[1], Limei Xu[1,3], Martin Ondráček[2], Pavel Jelínek[2,4†], Enge Wang[1,3†], and Ying Jiang[1,3†]

[1]*International Center for Quantum Materials, School of Physics, Peking University, Beijing 100871, P. R. China*

[2]*Institute of Physics, the Czech Academy of Sciences, Cukrovarnicka 10, 1862 53 Prague, Czech Republic*

[3]*Collaborative Innovation Center of Quantum Matter, Beijing 100871, P. R. China*

[4]*RCPTM, Palacky University, Šlechtitelů 27, 783 71, Olomouc, Czech Republic.*

**Scanning probe microscopy (SPM) has been extensively applied to probe interfacial water in many interdisciplinary fields but the disturbance of the probes on the hydrogen-bonding structure of water has remained an intractable problem. Here we report submolecular-resolution imaging of the water clusters on a NaCl(001) surface within the nearly non-invasive region by a qPlus-based noncontact atomic force microscopy. Comparison with theoretical simulations reveals that the key lies in probing the weak high-order electrostatic force between the quadrupole-like CO-terminated tip and the polar water molecules at**

---

[*] These authors contributed equally to this work.
[†] E-mail: jelinkep@fzu.cz (P.J.); egwang@pku.edu.cn (E.G.W.); yjiang@pku.edu.cn (Y.J.)

**large tip-water distances. This interaction allows the imaging and structural determination of the weakly bonded water clusters and even of their metastable states without inducing any disturbance. This work may open up new possibility of studying the intrinsic structure and electrostatics of ice or water on bulk insulating surfaces, ion hydration and biological water with atomic precision.**

INTRODUCTION

Water-solid interactions are of broad importance in many basic and applied fields, ranging from surface science to materials science and even bioscience ([1-4]). In particular, resolving the hydrogen-bonding (H-bonding) structure of interfacial water is crucial for understanding many extraordinary physical and chemical properties of water/solid interfaces. To date, scanning probe microscopy (SPM) including scanning tunneling microscopy (STM) ([5-16]) and atomic force microscopy (AFM) ([17-23]) has been an ideal tool to visualize the microscopic structure and dynamics of water at solid surfaces. However, an intrinsic problem of SPM is that all the probes inevitably induce perturbation to the fragile water structure, due to the excitation of the tunneling electrons and the tip-water interaction forces, especially under the close-imaging condition applied in order to achieve ultrahigh spatial resolution. This limitation makes SPM fall short compared with non-invasive spectroscopic methods such as optical spectroscopy, neutron scattering and nuclear magnetic resonance.

Recent advances in qPlus-based noncontact AFM (nc-AFM) show the ability to achieve superior resolution of aromatic molecules in real space using functionalized tips, such as identifying the chemical structure and intermolecular interaction (*24-28*), determining the bond order (*29*) and chemical-reaction products (*30*), imaging the charge distribution within a molecule (*31*), and even reveal the internal structure of metal clusters (*32*). Unfortunately, the atomic resolution of nc-AFM is only achieved at the very small tip-molecule separation where the short-range Pauli repulsion force is dominant *(24, 33)*. The tip-molecule interaction in this range is quite strong such that significant relaxation of the tip apex is induced (*33*). Considering that H bonds are much weaker than covalent bonds, the water structure may be easily disturbed at small tip heights (*21*). At large tip heights where only the long-range van der Waals and electrostatic forces are detectable, the resolution is usually quite poor. However, in contrary to the weakly polarized aromatic molecules, the water molecule has a strong internal dipole moment. Therefore, the imaging mechanism driven by the electrostatic force greatly relies on the detailed charge nature of the tip apex (*34, 35*).

Herein, we report the submolecular-resolution imaging of water nanoclusters on a Au-supported NaCl(001) film by probing the high-order electrostatic force using a qPlus-based nc-AFM. The AFM images of the water tetramers taken with a CO-terminated tip at large tip-water distance show prominent internal features, which resemble the electrostatic potential distribution within the cyclic tetramer. Comparison with the theoretical simulations reveals that such a high resolution originates from the electrostatic force acting between the quadrupole-like CO-tip and the strongly polar water molecules. In contrast, the results obtained with a monopole-like Cl-tip show much poorer resolution at large tip heights, arising from the different decay behaviors of the tip-water electrostatic interaction and the different charge distribution at the tip apex. Strikingly, the multipole electrostatic force between the CO-tip and water is rather weak, thus allowing precise structural determination of the weakly bonded water clusters and even their metastable states without inducing any disturbance.

**RESULTS AND DISCUSSION**

The experimental set-up is schematically shown in Fig. 1A, where the tip apex is functionalized with a CO molecule (See Methods). Water tetramers were constructed by assembling four individual $H_2O$ monomers on the NaCl(001) surface at 5 K. Our previous work reveals that each water molecule donates and accepts just one H bond resulting in a cyclic tetramer (Fig. 1B), whereas the other four free OH bonds point obliquely upward away from the surface (Fig. 1C) (*36*). In fact, the cyclic water tetramer may form two degenerate chiral H-bonded loops, which are respectively displayed in Fig. 1D and H, with the calculated Hartree potential superimposed. Fig. 1E-G and I-K are constant-height Δf images of the two degenerate tetramers acquired with the CO-tip at three different tip heights. At a large tip height, the two tetramers were imaged as four "ear-like" depressions with distinct chirality (Fig. 1E and I), which closely resemble their electrostatic potential (Fig. 1D and H). As the tip height decreased, the H-bonded loop was visualized as a bright square (Fig. 1F and J). When further approaching the tip, besides the sharpening of the square lines, contrast inversion was also observed at the center of the tetramer (Fig. 1G and K).

Interestingly, from Fig. 1F, G, J and K it is evident that the chiral contrast almost vanishes at small tip heights.

It is very unusual to obtain submolecular contrast (Fig. 1E and I) at large tip heights where the long-range force dominates the tip-water interaction. To understand the imaging mechanism, we used a molecular mechanics tip model (See Methods) to simulate the AFM images (Fig. 2A-D). We analyze the AFM contrast at different tip heights $z_1$, $z_2$, and $z_3$ as denoted in Fig. 2E. The simulated Δf images of an anticlockwise tetramer with the neutral tip model (Fig. 2A, $z_2$ and $z_3$) agree well with the experimental results at small tip heights (Fig. 1J and K). Detailed analysis (see fig. S1 for details) reveals that the sharp lines and the contrast inversion both result from the Pauli repulsion and the consequent lateral relaxation of the CO molecule at the tip apex, similar to previous studies of aromatic molecules(*29*, *37*). The sharp edges observed in AFM images should not be automatically related to presence of interatomic bonds, instead they represent ridges of the potential energy landscape experienced by the functionalized probe(*21, 33*).

However, the simulation for the neutral tip at the large tip height (Fig. 2A, $z_1$) fails to reproduce the internal chiral structure of tetramer (Fig. 1I). Fig. 2B-D compare simulated Δf images using monopole (s), dipole ($p_z$) and quadrupole ($d_{z^2}$) tip models, respectively (fig. S2). The simulated images with the monopole and dipole tips at the large tip height (Fig. 2B and C, $z_1$) show very little chirality. In contrast, the "ear-like" chiral features in Fig. 1I can be perfectly reproduced with the quadrupole tip (Fig. 2D, $z_1$), which also yields good agreement with the experimental images at the small tip heights (Fig. 2D, $z_2$ and $z_3$). We note that the simulation results are insensitive to the stiffness (k) of the tip, but greatly rely on the effective charge density (Q) (fig. S3). In fact, the quadrupole nature of the CO-tip can be verified from the plot of charge density difference calculated by density functional theory (DFT) (Fig. 2F). It is a result of charge redistribution between the adsorbed CO and metal tip(*35*).

The variation of the AFM contrast using different tip models can be understood from the analysis of calculated electrostatic forces acting between the sample and the given tip model. In Fig. 2G we plot xz-cut planes of vertical electrostatic force, which show significantly different shapes and decay behaviors for different charged tip models. Indeed, from the simulated electrostatic force curves over the water tetramer, we can see that the electrostatic force between the quadrupole tip and water decays much faster than the others as increasing the tip height (fig. S4 and Table S1). Such a difference in the decay behavior can be also seen from the experimental force curves acquired with CO-tip (quadrupole) and Cl-tip (monopole) (fig. S4 and Table S1). The long-range electrostatic force between the monopole tip and water only creates a large attractive background in the AFM images, thus hindering submolecular contrast.

Furthermore, we note that the lateral potential profile of CO-tip apex resembles well the "Mexican hat" wavelet (see fig. S5), which acts as an internal high-pass filter (actually Laplace filter, see also fig. S2). A tip with such kind of charge distribution can filter out the smoothly varying force components and becomes more sensitive to the atomic details. To the best of our knowledge, such a high-resolution image of electrostatic force has never been achieved for aromatic molecules with the CO-tip at far tip-sample distances. The main reason is that the water molecule has a much larger dipole moment than those molecules. In such a case, the multipole charge distribution of the CO-tip and the related electrostatic force should be taken into consideration to explain the improved resolution.

In order to verify the proposed imaging mechanism above, we functionalized the tip apex with a Cl atom. According to our previous DFT simulations (*15, 38*), the Cl atom is negatively charged with about 0.3-0.4 e when attached to the metal tip, acting as a monopole tip (see fig. S5). Fig. 3A and E display constant-height Δf images of the two degenerate tetramers recorded with a Cl-tip at a large tip height, displaying negligible chirality. This is consistent with the AFM simulation using the monopole tip (Fig. 2B, $z_1$), revealing the low sensitivity of monopole-like probe charges for high-resolution mapping of complex electrostatic fields. At smaller tip heights, the Δf images (Fig. 3B and F) show prominent sharp squares and "fork-like" features at the periphery (see the green arrows), also agreeing well with the simulation (Fig. 2B, $z_2$).

When using a smaller oscillation amplitude, the Δf images change remarkably, showing bright helical structures with distinct chirality (Fig. 3C and G), similar to the chiral depression observed with the CO-tip at the large tip height (Fig. 1E and I) (for the effect of oscillation amplitude, see fig. S6). From the simulations (Fig. 3D and H), we found that those chiral structures obtained with the Cl-tip arise from the pronounced tip relaxation at close tip-water distances, which is determined by the complex interplay between the Pauli and the electrostatic interaction (see fig. S1 and fig. S7). In contrast, we found that the quadrupole tip shows only negligible lateral relaxation at the large tip height where the submolecular electrostatic potential mapping is obtained (fig. S7), suggesting that the tip-water interaction force is very small such that the disturbance of the CO-tip on the water structure should be minimal in this range. This may open up the possibility of probing weakly bonded water clusters other than the rigid tetramers.

To confirm this possibility, we investigated fragile water structures such as dimers and trimers, which are very difficult to image with STM. Fig. 4A-C are the geometric structures, experimental and simulated Δf images of three water dimers at a large tip height, respectively. Similarly, we found that the depression features directly reflect the distribution of electrostatic potential in the water dimers (fig. S8). It is worthy to be noted that the crooked depressions in the AFM images are actually correlated with the position of the H atoms, which can help us identify the detailed configuration of various water clusters with unprecedented precision. It is striking that the AFM imaging can readily distinguish the subtle difference of the O-H tilting in the water dimers, whose energy barrier is as small as ~20 meV according to DFT calculations (fig. S9).

Water trimers are even more unstable than the dimers since they can have many metastable states, but we are still able to image the electrostatic potential of various water trimers with submolecular resolution (Fig. 4D-F and fig. S8). In combination with the simulations, their atomic configurations can be unambiguously determined. The calculated adsorption energies of those metastable water trimers are very close, with the smallest difference about only 10 meV, which is almost within the accuracy of DFT. The ability of discerning them suggests that our probe is indeed non-invasive. This technique is also applicable for more complicated water structures such as bilayer triple-tetramers (fig. S10). It is most surprising that the chirality of the middle tetramer in the triple-tetramer can be well resolved, although it is somewhat blocked by the higher bridging water molecules.

It is worthy to recall that the Cl-tip can also obtain submolecular-resolution imaging of the electrostatic potential of water tetramer by using small oscillation amplitudes (Fig. 3C and G). However, such a resolution is only achieved at small tip-water separation where the electrostatic and Pauli force becomes strong enough to induce significant relaxation of the tip apex. Any attempts to enter into this region can easily disturb the weakly bonded water clusters such as the water dimers, trimers and bilayer ice clusters (fig. S11). Therefore, the high-order electrostatic force between the CO-tip and the water is critical since it yields submolecular resolution at large tip-water separations where the electrostatic force and other force components are still rather weak, thus avoiding the disturbance of the tip on the water molecules.

**CONCLUSION**

In summary, we have achieved non-perturbative imaging of weakly bonded water clusters, which defeats the longstanding limitation in the SPM studies of water. The submolecular-resolution AFM images of water obtained by CO-tip not only provide the spatial information of electrostatics, but also allow us to determine the detailed H-bonding structure including the position of the H atoms, which is crucial for the understanding of H-bonding interaction and dynamics of water. Furthermore, those results shed new light on the mechanism of high-resolution AFM images, highlighting the key roles of the complex charge distribution of the tip apex in the imaging of the polar molecules. This work may open up a new avenue for studying ice or water on bulk insulating surfaces, ion hydration, and biological water with atomic precision.

## MATERIALS AND METHODS

**STM/AFM experiments.** All the experiments were performed with a combined nc-AFM/STM system (Createc, Germany) at 5 K using a qPlus sensor equipped with a W tip (spring constant $k_0 \approx 1800$ N/m, resonance frequency $f_0 = 23.7$ kHz, and quality factor $Q \approx 80000$). The NaCl(001) bilayer film was obtained by thermally evaporating NaCl crystals onto a clean Au(111) surface at room temperature. The ultrapure $H_2O$ (Sigma Aldrich, deuterium-depleted) was used and further purified under vacuum by several freeze-and-pump cycles to remove remaining impurities. The $H_2O$ molecules were dosed in situ onto the sample surface at 5 K through a dosing tube. All of the frequency shift ($\Delta f$) images were obtained in constant-height mode at 5 K with Cl- or CO-terminated tips. The preparation of the Cl-tip was the same as in Ref. (*15*). The CO-tip was obtained by positioning the tip over a CO molecule at a set point of 100 mV and 20 pA, followed by increasing the bias voltage to 200 mV. The controllable manipulation of water monomers to construct water tetramers was achieved with the Cl-terminated tip at the set point: V=10 mV, I=150 pA.

**Simulations of AFM images.** The Δf images were simulated with a molecular mechanics model including the electrostatic force, based on the methods described in Refs. (*37*) and (*33*). We used the following parameters of the flexible probe-particle tip model: the effective stiffness k = 0.5 N/m and effective atomic radius $R_c$ = 1.66 Å. In order to extract the effect of electrostatics more clearly and to make z-distance directly comparable, we used the same stiffness and atomic radius to simulate AFM images acquired with CO and Cl-terminated tips. Noteworthy, the simulated Δf images using different atomic radius of the probe particle to mimick CO ($R_c$=1.66 Å) and Cl ($R_c$=1.95 Å) tip-apex models with the same effective charges display essentially the same features. The input electrostatic potentials of water tetramer (using previously optimized atomic structure from Ref. (*36*)) and other water clusters were obtained by DFT calculation using the VASP code with a plane-wave cutoff 600 eV and 550 eV, respectively. Parameters of Lennard Jones pairwise potentials for all elements are listed in Table S2.

**DFT calculations.** DFT calculations were performed using the Vienna ab-initio simulation package (VASP; Ref. (*39*), Projector augmented wave method (PAW; Ref. (*40*)) with PBE functional (*41*) were used. Van der Waals corrections for dispersion forces were considered using the van der Waals density functional scheme with the optB88-vdW method (*42*). Similar to Ref. (*36*), we used a bilayer NaCl(001) slab separated by a vacuum thicker than 20 Å and the bottom layer of the NaCl was fixed with a bulk lattice constant of 5.665 Å. Supercells with Monkhorst-Pack k-point meshes of spacing denser than $2\pi \times 0.042 Å^{-1}$ and a plane-wave cutoff 550 eV were used. The geometry optimizations were run with the energy criterion of 5×10$^{-5}$ eV and the adsorption energy was calculated by subtracting the total energy of the nH$_2$O/NaCl(001) structure from the sum of the energies of the relaxed bare NaCl(001) substrate and n isolated water molecules in gas phase:

$$E_{ads} = E[(NaCl(001))_{relaxed} + n \times E[(H_2O)_{gas}] - E[(NaCl(001) + nH_2O)_{relaxed}]$$

Energy barrier of water dimer was determined using the climbing image nudged elastic band (cNEB) method (*43*) with the force criterion of 0.02 eV/Å.

**Acknowledgements**

This work was supported by the National Key R&D Program under Grant No. 2016YFA0300901 and 2016YFA0300903, the National Natural Science Foundation of China under Grant No. 91321309, 11290162/A040106. Y. J. acknowledges support

by National Program for Support of Top-notch Young Professionals. P.H. and P.J. acknowledge support of GAČR project No. 14-16963J. J.G. acknowledges support from the National Postdoctoral Program for Innovative Talents. J.P. acknowledges support from the Weng Hongwu Original Research Foundation under Grant No. WHW201502.

**Author contributions**

Y.J. and E.G.W. designed and supervised the project. J.P., J.G. and R.M. performed the STM/AFM measurements. P.H., M.O. and P.J. carried out the theoretical simulations of the AFM images in collaboration with D.C. and B.W.. D.C. and L.X. which performed the total energy DFT calculations. J.P., J.G., P.H., P.J., D.C., R.M., B.W., L.X., E.W., and Y.J. analyzed the data. Y.J., J.P. and J.G. wrote the manuscript with P.H., P.J. and E.G.W. The manuscript reflects the contributions of all authors.

**Competing financial interests**

The authors declare no competing financial interests.

**Figure Captions:**

**Fig. 1. Experimental set-up and AFM images of two degenerate water tetramers with a CO-terminated tip.** (**A**) Schematic of a qPlus-based nc-AFM with a CO-tip. The cantilever oscillates at an amplitude of A and the tip-sample force induced frequency shift of the cantilever from its natural resonance frequency ($f_0$) is Δf. (**B and C**) Top and side view of the water tetramer adsorbed on the NaCl(001) surface, respectively. H, O, Cl, Na atoms are denoted as white, red, green and purple spheres, respectively. (**D-G**) and (**H-K**) Water tetramers with clockwise and anticlockwise H-bonded loops, respectively. (D) and (H) Calculated electrostatic potential map of the water tetramers in a plane 60 pm above the outermost H atom. (E and I), (F and J), (G and K) Experimental Δf images recorded at the tip heights of 100 pm, 10 pm, -40 pm, respectively. The tip height is referenced to the STM set point on the NaCl surface (100 mV, 50 pA). The oscillation amplitude is 100 pm. The size of the images is 1.2 nm× 1.2 nm.

**Fig. 2. The role of electrostatics in the high-resolution AFM imaging of a water tetramer.** (**A-D**) Simulated AFM images of a water tetramer with neutral, s, $p_z$ and $d_{z^2}$ tip models, respectively (k = 0.5 N/m, Q= -0.2e). The first, second and third rows correspond to the images acquired at the tip heights of about $z_1$=7.9 Å, $z_2$=6.8 Å and $z_3$=6.4 Å, respectively. For a better comparison, we had chosen similar simulation images by subtracting a small offset of tip height between different tips. The tip height is defined as the distance between the outmost metal atom of the tip and the upward H atom of the water tetramer. The oscillation amplitude of all the simulated

Δf images is 100 pm. The size of the images is 1.2 nm× 1.2 nm. (**E**) Simulated force curve of the water tetramer taken with the $d_{z^2}$ tip, where the three tip heights ($z_1$, $z_2$ and $z_3$) are denoted. (**F**) Charge distribution of the CO-tip from DFT calculations. (**G**) Maps of calculated vertical electrostatic forces between the sample and different tip models (s, $p_z$ and $d_{z^2}$) computed by convolution of Hartree potential of sample and model charge distribution on the tip (*37*).

**Fig. 3. Experimental and simulated AFM images of two degenerate water tetramers with a Cl-terminated tip.** (**A-C**) and (**E-G**) Experimental Δf images of the water tetramers with clockwise and anticlockwise hydrogen-bonded loops, respectively. The tip heights are 30 pm (A and E), -120 pm (B and F), -120 pm (C and G). The oscillation amplitudes are 40 pm (A and E), 100 pm (B and F), 40 pm (C and G). The "fork-like" features are denoted by two green arrows in (B) and (F). (**D**) and (**H**) Simulated Δf images with the oscillation amplitudes of 40 pm, which were obtained with a monopole (s) tip (k = 0.5 N/m, Q = -0.25e). The size of the images is 1.2 nm× 1.2 nm.

**Fig. 4. Submolecular-resolution AFM images of weakly bonded water clusters with a CO-tip.** (**A-C**) Geometric structures, experimental and simulated Δf images of two water dimers, respectively. The tip height of (B) is 100 pm, 100 pm and 130 pm (from left to right), respectively. (**D-F**) Geometric structures, experimental and simulated Δf images of three water trimers, respectively. The tip height of (E) is 130 pm, 130 pm and 110 pm (from left to right), respectively. All the oscillation amplitudes of experimental and simulated images are 100 pm. All the simulations

were done with a quadrupole ($d_{z^2}$) tip (k= 0.5 N/m, Q= -0.2e). The size of the images: 1.2 nm× 1.2 nm.

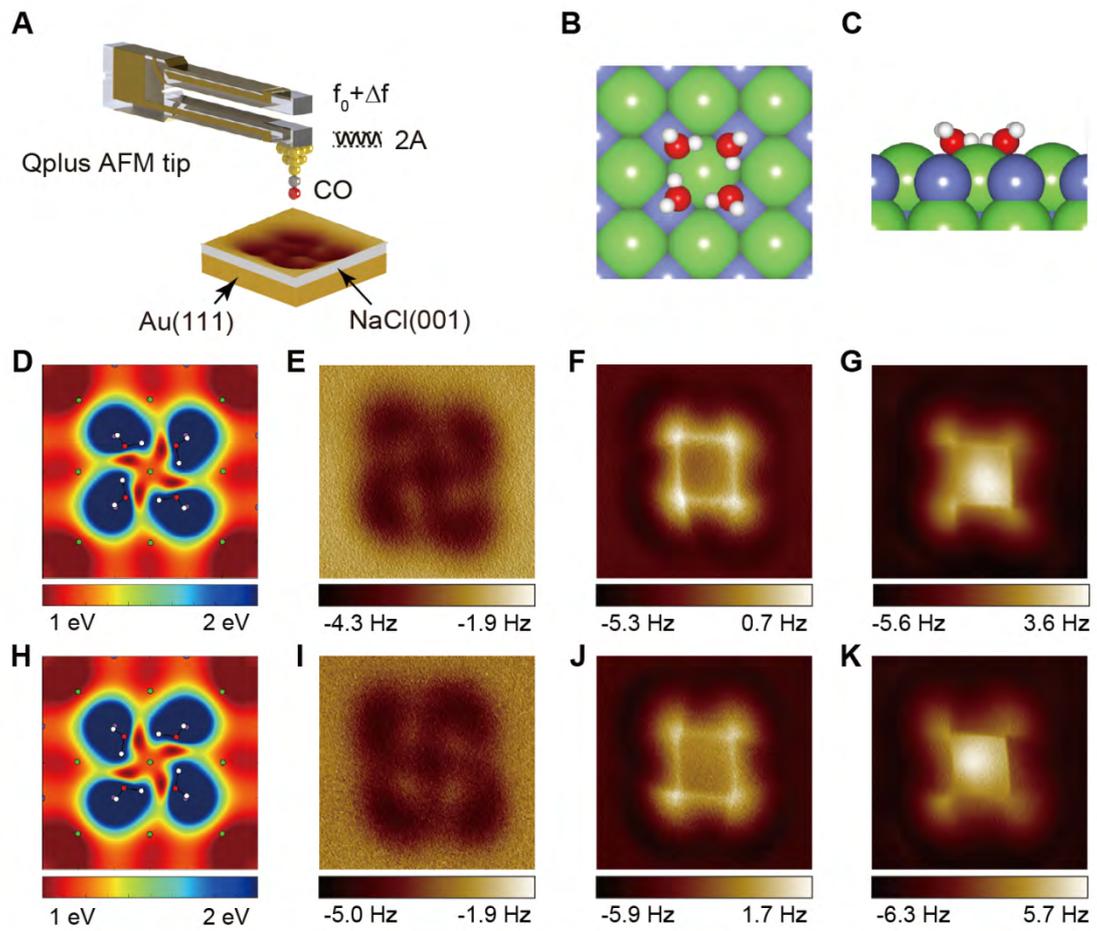

Figure 1

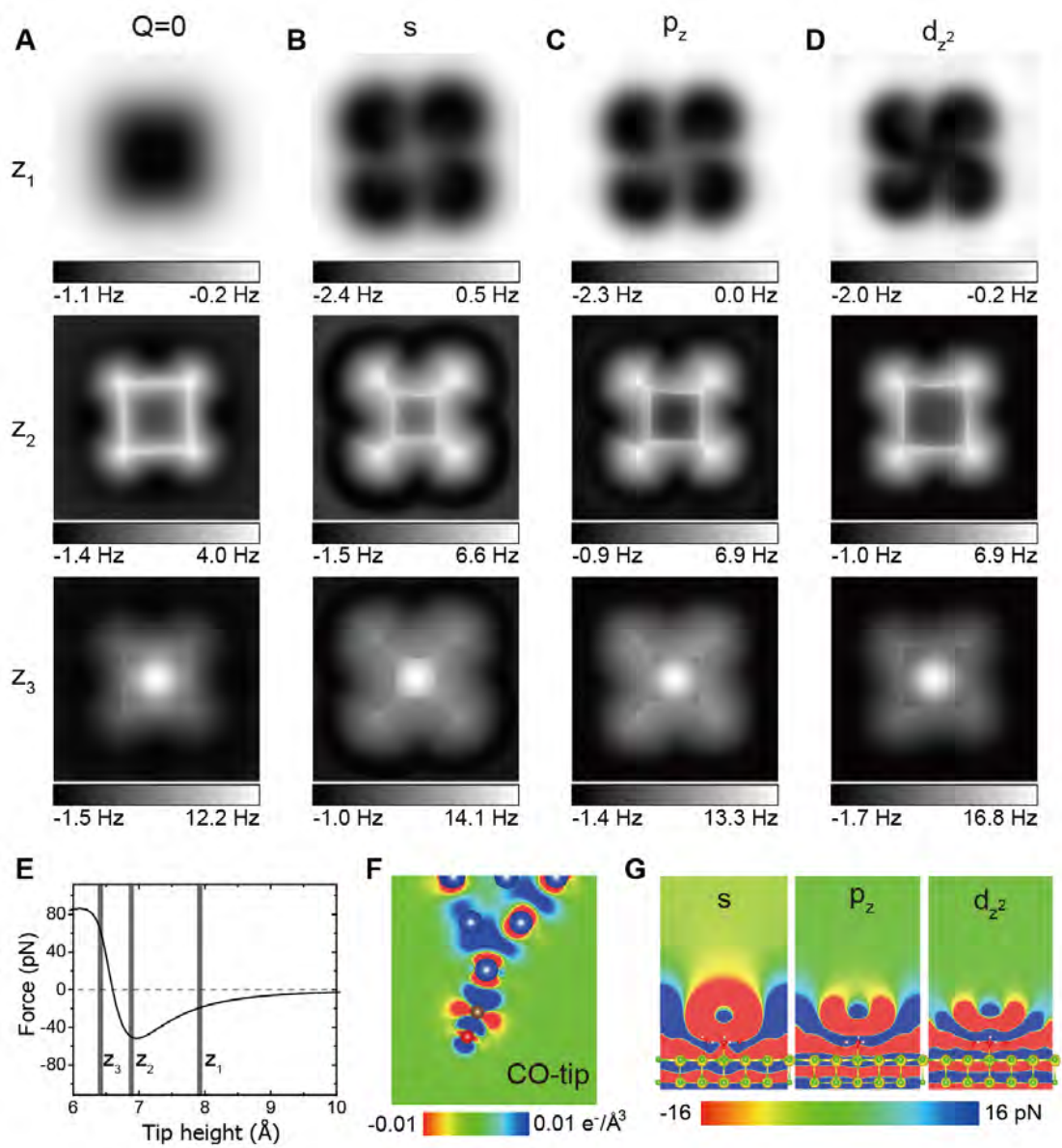

**Figure 2**

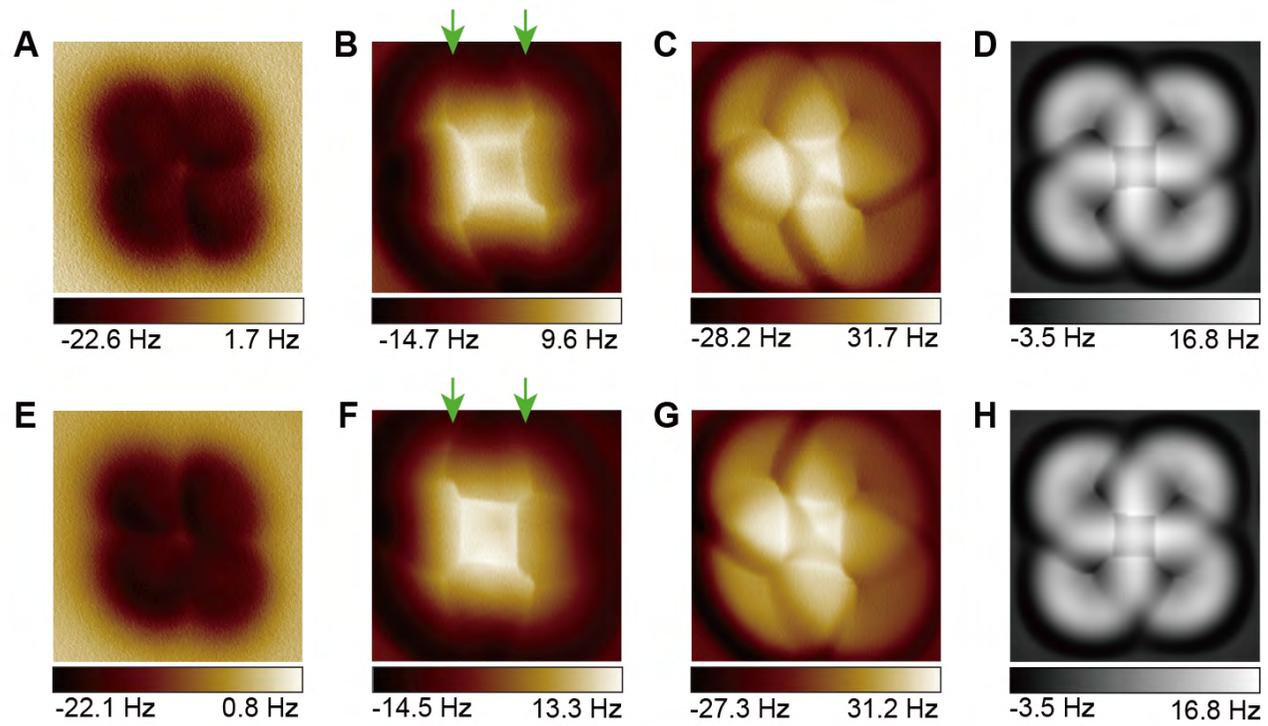

**Figure 3**

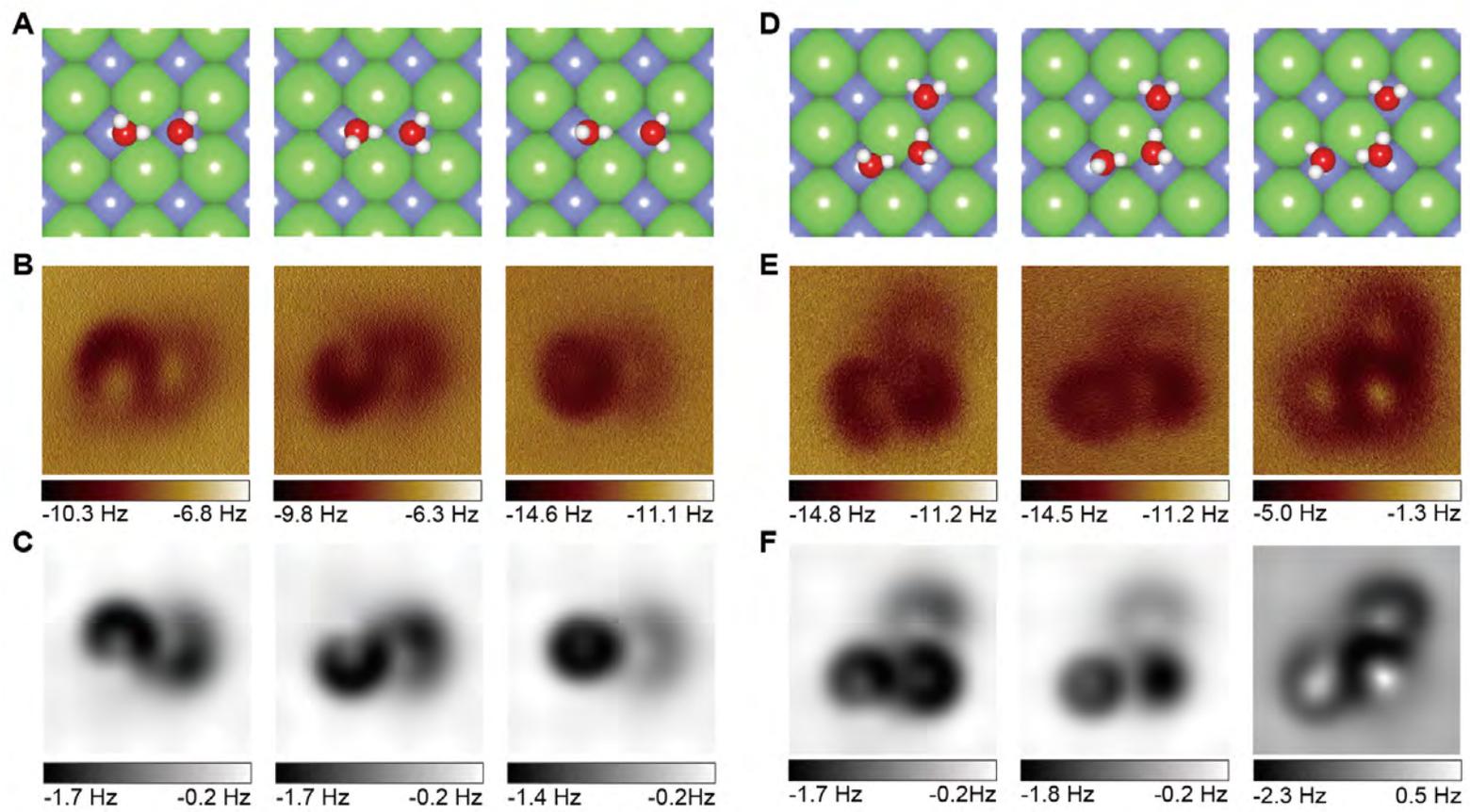

**Figure 4**

Supporting Online Material for

# Submolecular-resolution non-invasive imaging of interfacial water with atomic force microscopy

Jinbo Peng[1*], Jing Guo[1*], Prokop Hapala[2*], Duanyun Cao[1], Runze Ma[1], Bowei Cheng[1], Limei Xu[1,3], Martin Ondráček[2], Pavel Jelínek[2,4†], Enge Wang[1,3†], and Ying Jiang[1,3†]

[1]International Center for Quantum Materials, School of Physics, Peking University, Beijing 100871, P. R. China

[2]Institute of Physics, the Czech Academy of Sciences, Cukrovarnicka 10, 1862 53 Prague, Czech Republic

[3]Collaborative Innovation Center of Quantum Matter, Beijing 100871, P. R. China

[4]RCPTM, Palacky University, Šlechtitelů 27, 783 71, Olomouc, Czech Republic.

## Contents:

---

[*] These authors contributed equally to this work.
[†] E-mail: jelinkep@fzu.cz (P.J.); egwang@pku.edu.cn (E.G.W.); yjiang@pku.edu.cn (Y.J.)

# I. Submolecular contrasts in Δf images at small tip heights

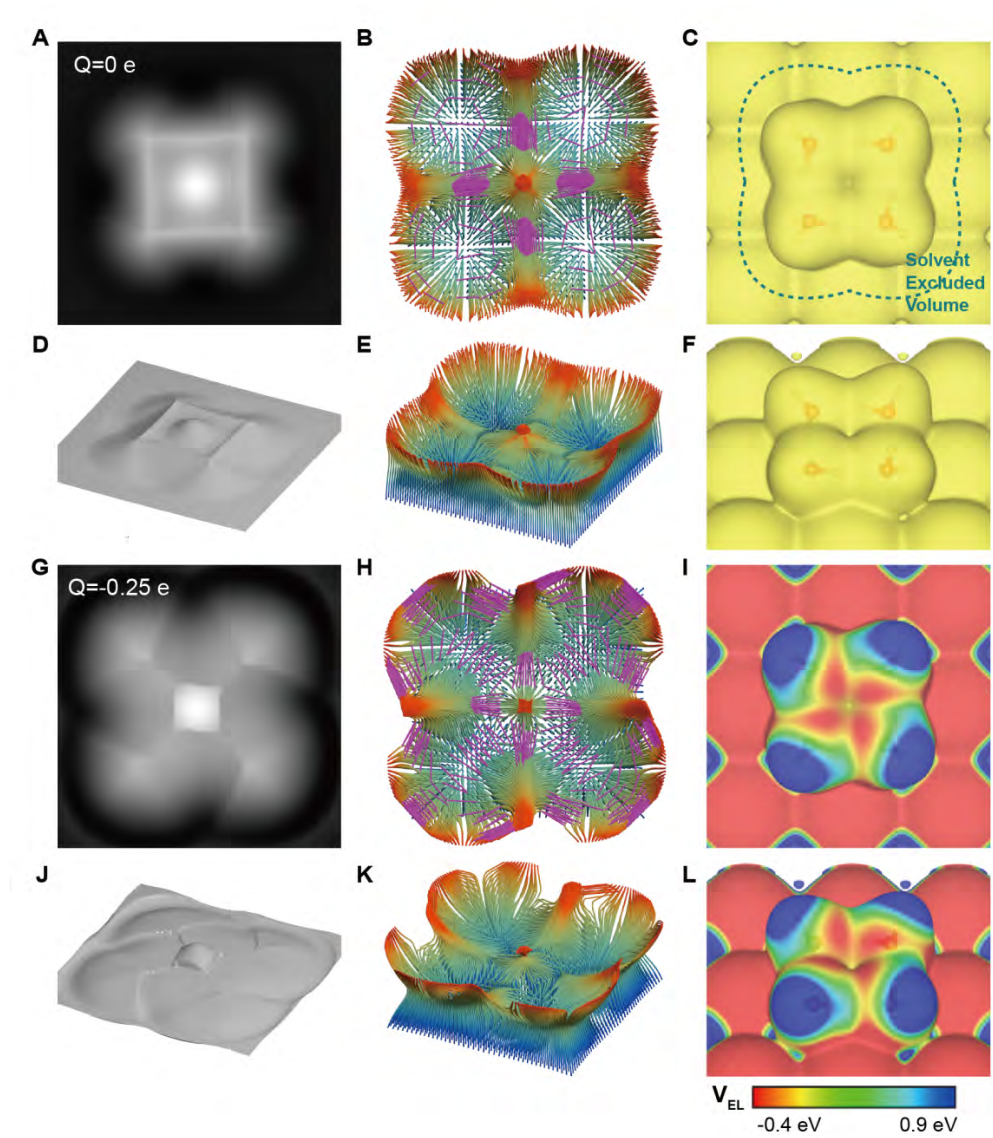

**fig. S1. Simulated Δf images, probe-particle trajectories, and Pauli and electrostatic potentials for water tetramer.** (**A-F**) Neutral tip. (**G-L**) Negatively charged monopole tip (Q=-0.25e). (**A** and **G**) Simulated Δf images. (**D** and **J**) 3D relief generated from the simulated Δf image. (**B**,**H**) and (**E**,**K**) Top and 3D view of probe-particle trajectories when approaching the tip to the surface (from blue to red). Purple lines denote branching of trajectories. More specifically, they were plotted when two neighboring trajectories diverge by more than 0.4 Å. (**C** and **F**) Isosurfaces of the total electron density which probe-particle cannot penetrate due to the Pauli repulsion. (**I** and **L**) Electrostatic potential mapped on top of the electron density isosurface (C and F). The presence of the strong electrostatic fields varies

trajectories of charged probe particle. The trajectory modification introduces additional branching points (H and K), giving rise to the fork-like features in the Δf images (G and J).

The sharp lines in the Δf images emerge from branching of probe-particle trajectories over saddle points of the total tip-sample interaction potential at small tip-water separations as discussed in (*33*). In the case of non-planar and strongly polarized system, such as water clusters, it leads to even more intriguing and unintuitive results, which deserve detailed discussion. The total interaction potential between the functionalized tip and the water molecules adsorbed on surface consists of Pauli repulsion, London dispersion and electrostatic interaction.

In the case of CO-tip, the image contrast can be fully understood by simulations (fig. S1A-F) that consider just the former two components of the potential (Pauli repulsion and London dispersion). This assumption can be justified by a small charge presented on the CO-tip (see fig. S5). Characteristic sharp square lines appear between the upward H atoms as a result of the saddles in the Pauli repulsion, which are visible also in a contour of the total electron density of the cluster (see fig. S1C ad F). Due to the finite van der Waals radius of the probe particle (see fig. S5), it moves around on slightly larger surface as described by the concept of "solvent excluded volume" introduced in biochemistry (*44*). The potential saddles leads to branching of the probe particle trajectories (fig. S1B and E), which gives rise to the sharp square in the Δf images (fig. S1A and D). The center of the sharp square exhibits contrast inversion at very close tip-sample distance (see fig. 1G and K) as the probe particle is locked in the center of the square and further relaxation is prevented.

On the contrary, the image contrast acquired with the Cl-tip is strongly affected by the electrostatic field of the water cluster, leading to very different features at small tip-sample distance, which can be also reproduced by our simulations using a monopole tip (see fig. S1G and J, fig. 3D and H). Based on these simulations, we can rationalize the origin of two main differences compared with the CO-tip case: (i) the shrinking of central square and (ii) appearance of additional fork-like features at the periphery (large amplitude, fig. 3B and F) and chiral ear-like rings (small amplitude, fig. 3C and G). All these features can be ultimately tracked down to a map of electrostatic potential (fig. S1I and L) overlaid on top of a contour of total electron density (or Pauli repulsion) along which the probe particle slides upon tip approaching. In the case of Cl-tip, the presence of the electrostatic field above the water tetramer makes the relaxation of probe particle more complicated. The probe particle (Cl ion) is repelled from negatively charged center toward positively charged H atoms, but then it suddenly slips off due to the Pauli repulsion over protruding H atoms and the restoring spring force of the tip. This sudden slip-off leads to additional branching of the probe particle trajectories ultimately manifested as discontinuity of Δf signal measured on different sides of branching line. Thus, it gives rise to the sharp fork-like features and the chiral ear-like rings in the Δf images. The exact position of branching lines is very sensitive to the detailed force balance between electrostatic and other forces (Pauli repulsion, restoring spring force). Therefore, the Δf images obtained with Cl-tip at small tip heights contain some information of the electrostatic field, which is strongly entangled with other force fields.

## II. The definition of charged tip models

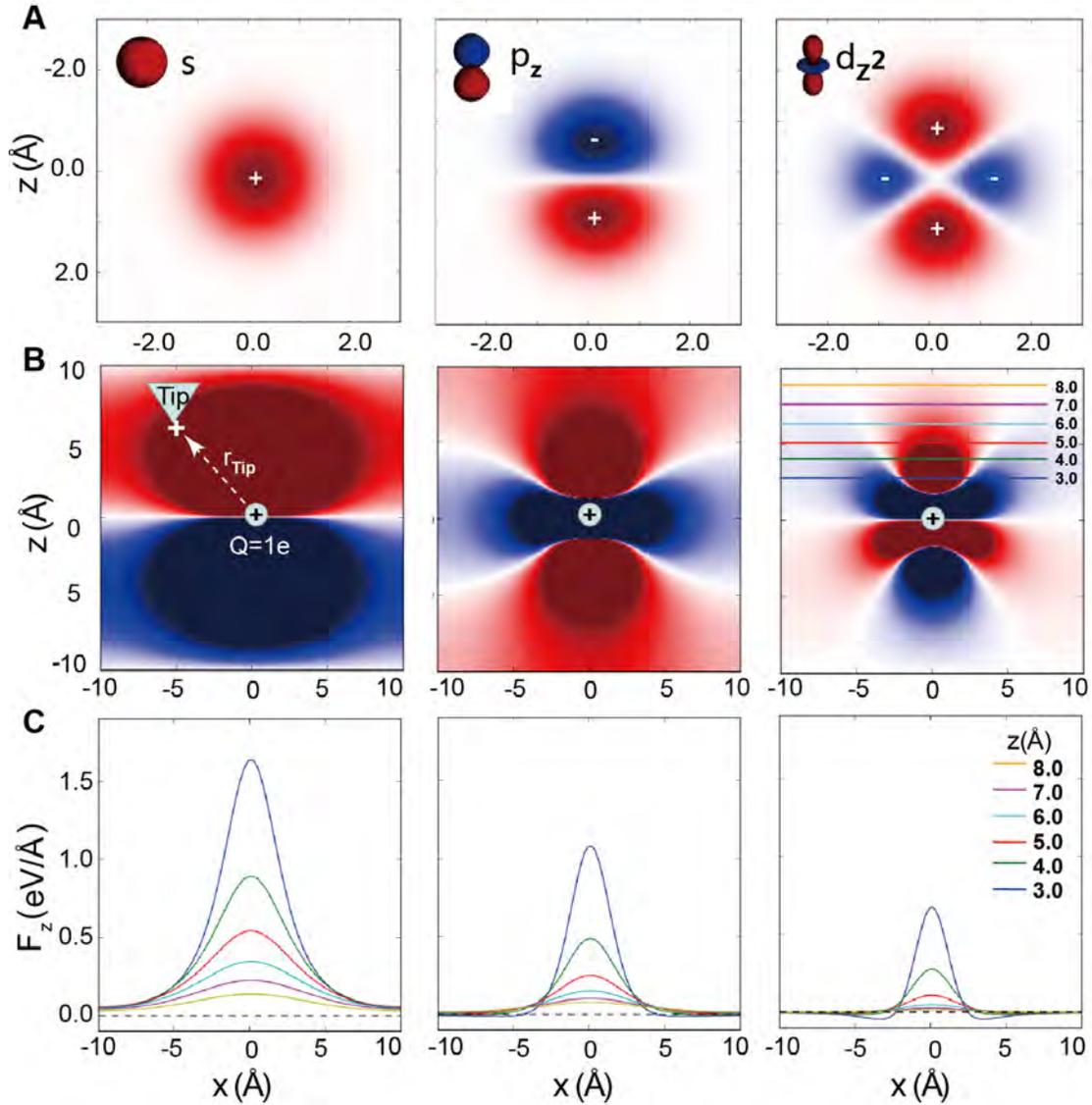

**fig. S2. Charge distribution and electrostatic force of different charged tip models.** (A) xz-cut of the charge distribution of monopole (s-like), dipole ($p_z$) and quadrupole ($d_{z^2}$), respectively. Detailed description of formulas is given below. We used the smearing width (effective radius) $\sigma = 0.7$ in all simulations presented in this work. (B) xz-cut of the vertical electrostatic force ($F_z$) between a point charge (Q=1e, as shown in the center) and s, $p_z$, $d_{z^2}$ tips which move around the point charge. (C) Line profile along x-axis of the vertical electrostatic force $F_z$ at different z distances (as indicated in B).

Here, we give specific formulas that define charge distribution on the tip for the multipole tip models. We have discussed three different models in our present paper: monopole (s), dipole ($p_z$) and quadrupole ($d_{z^2}$) (see fig. S2A). A general formula for the spatial distribution of charge density corresponding to a multipolar tip can be written as

$$\rho(x,y,z) = Q R_\sigma(r) \phi(x,y,z),$$

where

$$r = \sqrt{x^2 + y^2 + z^2},$$

the function $R_\sigma$ is a normalized three-dimensional Gaussian that defines the radial part of the charge density:

$$R_\sigma(r) = \frac{e^{-\frac{r^2}{2\sigma^2}}}{(\sqrt{2\pi}\sigma)^3}$$

and the angular part $\phi(x,y,z)$, specific for the multipole in question, is

$$\phi_s(x,y,z) = 1,$$

$$\phi_{p_z}(x,y,z) = z/\sigma,$$

$$\phi_{d_{z^2}}(x,y,z) = \frac{2z^2 - x^2 - y^2}{4\sigma^2}.$$

There are two parameters to be chosen for each of these model distributions: The smearing width (effective radius) $\sigma$ and an overall multiplicative factor $Q$. In this paper, we adopt value of $\sigma = 0.7$. The normalization of the functions $R_\sigma$, $\phi_s$, $\phi_{p_z}$, and $\phi_{d_{z^2}}$ was chosen so that

$$\int \rho_s(x,y,z)\, dx\, dy\, dz = Q,$$

$$\int z \rho_{p_z}(x,y,z)\, dx\, dy\, dz = Q\sigma,$$

$$\int z^2 \rho_{d_{z^2}}(x,y,z)\, dx\, dy\, dz = Q\sigma^2.$$

This choice gives a straightforward interpretation of the factor $Q$. For a monopole, it is simply the total charge. For a dipole and quadrupole, $Q\sigma$ and $Q\sigma^2$, respectively, give its magnitude.

From the definition of charged tip models, the charge density of a quadrupole is

$$\rho_{d_{z^2}}(x,y,z) = Q \frac{(2z^2 - x^2 - y^2)}{4\sigma^2} R_\sigma(r)$$

Equivalently, it can be written as

$$\rho_{d_{z^2}}(x,y,z) = \frac{Q\sigma^2}{4}\left(2\frac{\partial^2}{\partial z^2} - \frac{\partial^2}{\partial x^2} - \frac{\partial^2}{\partial y^2}\right)R_\sigma(r).$$

Thus, it can be considered as a linear combination of the 1D Laplace filter in the z direction and the 2D Laplace filter in the xy plane. Since the Laplace filter tends to emphasize the local changes of the electrostatic potential, enhanced spatial resolution is expected with a $d_{z^2}$ tip.

Such an effect can be seen very clearly in electrostatic force ($F_z$) between a point charge (as a test) and different tips (fig. S2B). From the x-profile of $F_z$ (fig. S2C), it is obvious that the peak width at half height with a $d_{z^2}$ tip is much smaller than that with an s tip or a $p_z$ tip. Besides, a "Mexican hat" shape can be seen at close distance (z=3 Å), which is also consistent with the DFT calculations in fig. S5. Therefore, the $d_{z^2}$ tip does show higher spatial resolution compared with the s tip and $p_z$ tip.

## III. Effect of the stiffness (k) and charge (Q) on the simulated AFM images

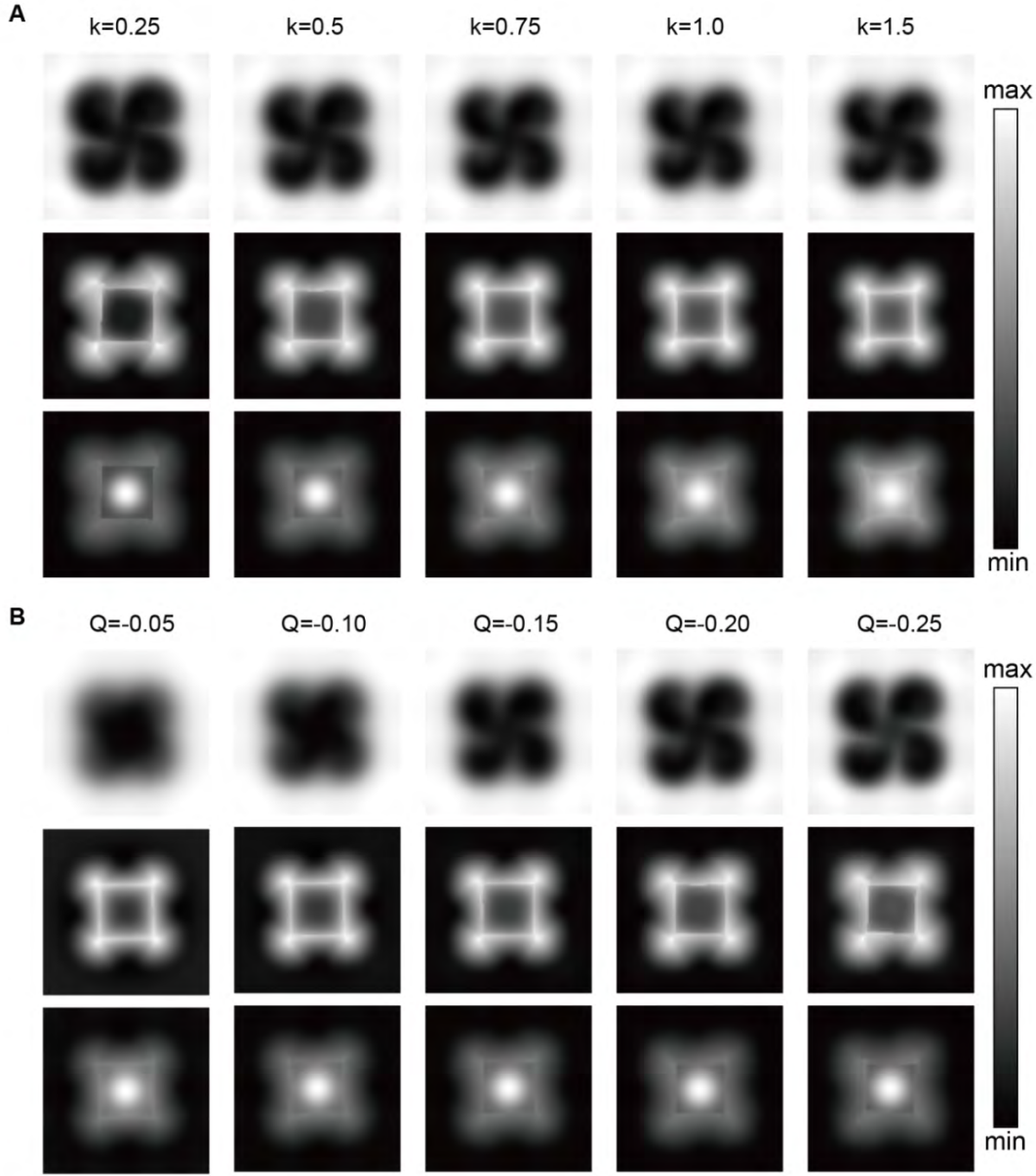

**fig. S3. The effect of the stiffness (k) and charge (Q) on the simulated AFM images of a water tetramer with a quadrupole ($d_{z^2}$) tip at different tip heights.** (**A**) The effect of the stiffness (k) on the simulated Δf images. (**B**) The effect of the charge (Q) on the simulated Δf images. The tip heights of the first, second and third rows are about 7.8 Å, 6.7 Å and 6.2 Å, respectively. For a better comparison, we had chosen similar simulation images by subtracting a small offset of tip height for different tips. The tip height in simulations is defined as the distance between the tip apex and the outmost H atom of water tetramer. All

the oscillation amplitudes are 100 pm. The size of the images is 1.2 nm× 1.2 nm.

As the stiffness (k) of the tip increases from 0.25 N/m to 1.5 N/m, the main features of AFM images do not change too much except for a slight distortion of the square at small tip height (fig. S3A, bottom), indicating the robustness of our simulation model. When the effective charge Q varies from -0.05e to -0.25e, the AFM images at the large tip height show an improved spatial resolution (fig. S3B, top), suggesting that the electrostatic force plays a key role in the AFM imaging of the tetramer. The simulated images with Q from -0.15e to -0.2e match the experimental results the best (fig. 1E and I). At small tip heights, the effective charge has little effect on the contrast of the images due to the dominant role of Pauli repulsion (fig. S3B, middle and bottom). We notice that the sharp square shrinks a lot with larger Q due to the lateral relaxation of the probe particle induced by the electrostatic force.

**IV. Decay length of the force curves with different tips**

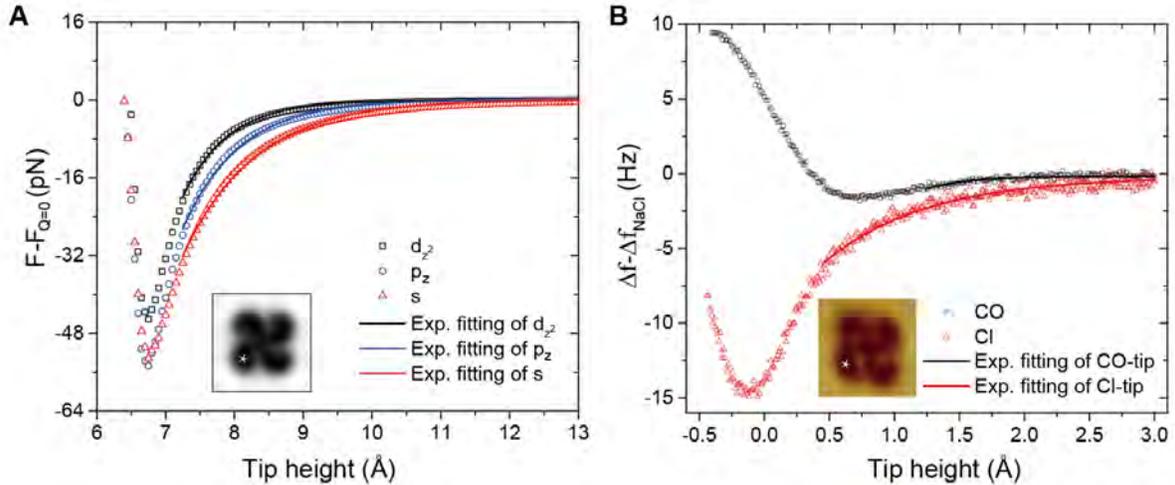

**fig. S4. The force curves between the water tetramer and different tips.** (**A**) The calculated force curves with s, $p_z$ and $d_{z^2}$ tips after subtraction of the force with a neutral tip. The tip position is indicated with a star in the inset. The solid lines are the corresponding exponential fittings of the curves within the range of tip height from 7.2 Å to 13 Å. The tip height is defined as the distance between the tip apex and the outmost H atom of water tetramer. A high simulation cell (4 nm) was used here in order to accommodate realistic decay of electrostatic field. (**B**) The frequency shift Δf measured above the water tetramer (as indicated with a star in the inset) with CO- and Cl-tips after removing the contribution from the NaCl substrate. The solid lines are the corresponding exponential fittings of the curves within the range of tip height from 0.45 Å (Cl-tip) or 1.25 Å (CO-tip) to 3 Å. The tip height is with respect to the set point of 100 mV and 50 pA on NaCl. The decay lengths of different tips are summarized in Table S1.

To extract the contribution of electrostatic force, we plotted the calculated force curves with s, $p_z$ and $d_{z^2}$ tips after subtraction of the force with a neutral tip (fig. S4A). Approximatively, we used an exponential fitting to obtain the decay length of the electrostatic force between the tetramer and different tips. To avoid the effect of tip relaxation at short tip-water separation, only the data points at large tip heights were fitted. The decay length of the $d_{z^2}$ tip is the smallest, as shown in Table S1. Similarly, we exponentially fitted the experimental Δf curves with CO- and Cl-tips after removing the contribution from the NaCl substrate (fig. S4B). We found that the decay length with the Cl-tip is more than two times larger than that with the CO-tip (Table S1), indicating the short-range nature of the high-order electrostatic force between the CO-tip and the water molecules.

## V. Electrostatic field of Cl-tip vs CO-tip

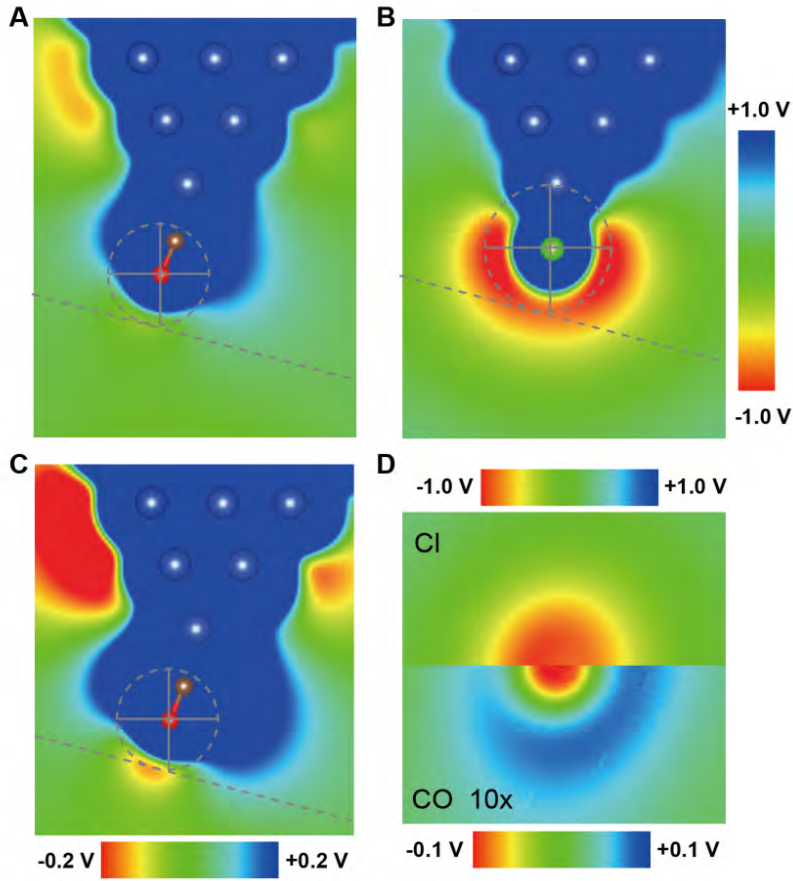

**fig. S5. Electrostatic field of Cl-tip vs CO-tip.** (**A** and **B**) xz-cut of Hartree potential of CO (A) and Cl (B) functionalized tip plotted in the same range (+/- 1.0V ) obtained from the total energy DFT simulations. Gray dotted circle denotes the van der Waals radius of the apex atom, Cl and O, respectively. The dark blue area around the tip atoms is due to unscreened potential of nuclei. Only the potential outside the vdW radius is relevant. (**C**) Hartree potential of the CO-tip plotted in a finer range +/-0.2V revealing a small negative cup below the oxygen atom. (**D**) Top view of the electrostatic potential (cut planes along the gray dotted lines shown in (A) and (B)) comparing Cl- and CO-terminated tips.

For comparison, we mapped the electrostatic field distribution of Cl-tip and CO-tip by DFT calculations (fig. S5A, B and C). As shown in fig. S5D, although the Cl-tip (upper half) has much stronger electrostatic field, the quadrupole-like CO-tip (lower half) has a highly localized negative potential at the CO apex showing a "Mexican hat" wavelet-like profile, which is quite similar to the Laplacian of Gaussian function. Thus, the CO-tip indeed behaves as a high-pass filter which can further enhance the spatial resolution by removing the slowly changed background. All these features agree quite well with that of the $d_{z^2}$ tip model (see fig. S2).

## VI. Effect of the oscillation amplitude on Δf images

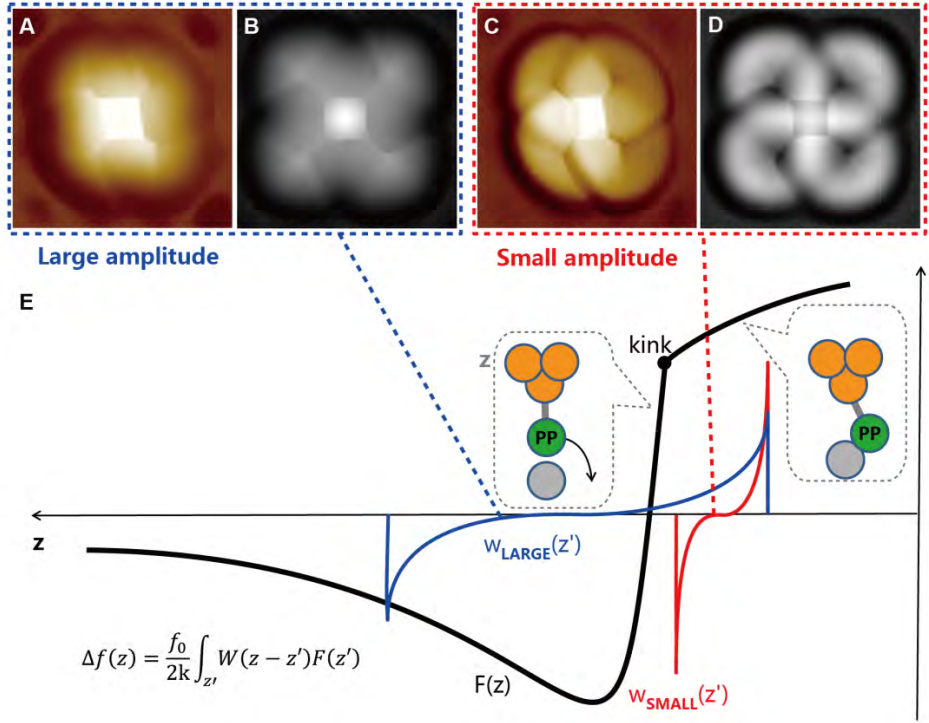

**fig. S6. Contrast variation in Δf images with oscillation amplitude.** (**A** and **B**) Experimental (A) and simulated (B) Δf images obtained with large oscillation amplitude. (**C** and **D**) Experimental (C) and simulated (D) Δf image obtained with small oscillation amplitude. (B) and (D) were obtained from the same simulated force data, using just different amplitude parameters in force-to-Δf conversion procedure. (**E**) Schematic diagram showing force vs. distance curve and weighting function $w$ for small (red) and large (blue) oscillation amplitude (*45*). The kink of the force curve is due to sudden lateral relaxation of the probe particle when lateral component of Pauli repulsion overcomes restoring spring force.

From comparison of AFM images acquired with the Cl-tip using large (fig. S6A and B) and small amplitude (fig. S6C and D), it is evident that the small-amplitude regime is much more sensitive to the chiral shape of the electrostatic potential. In general, the Δf signal results from a weighted convolution of the force over a range of the oscillation amplitude (*45*). In the case of large oscillation amplitude, the probe spends large part of the oscillation period at tip-sample distances, where the chirality of the electrostatic potential is almost negligible (fig. S6E, blue curve). In addition, the electrostatic potential changes significantly at the very close distance, having a non-trivial 3D chiral character. In the limit of the small amplitude, the frequency shift is proportional to derivative of force along z-distance (fig. S6E, red curve). Therefore, the non-trivial 3D character of the electrostatic potential induces a significant impact on the frequency shift when small amplitude is used. This chirality is further enhanced by a contrast inversion of the sharp features in AFM images (fig. S6A-D), which is caused by a sudden lowering of the slope of the force curve when the probe particle is deflected laterally (fig. S6E).

**VII. The relaxation of different tip apex in the AFM imaging**

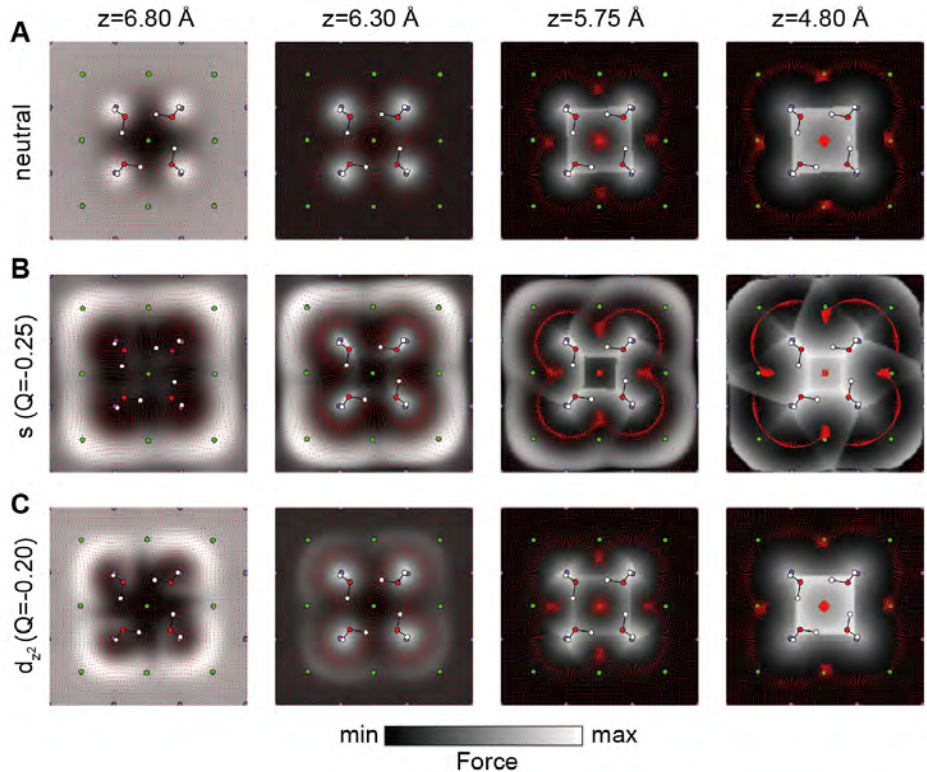

**fig. S7. The relaxation of different tip apex in the AFM imaging.** (A-C) Two-dimensional maps of the calculated vertical deflection (grey-scale background) and lateral relaxation (red

dots) of the probe particle, respectively. The vertical deflection is related to the vertical force by Hooks law. The schematic tetramer structure and the underneath NaCl lattice are superimposed in the maps. H, O, Na, Cl atoms are denoted as white, red, purple and green spheres, respectively. (A), (B) and (C) were obtained with neutral (Q = 0), s (Q = -0.25e) and $d_{z^2}$ (Q = -0.2e) tips, respectively. The stiffness (k) of the tips is 0.5 N/m. The definition of the tip height z is the same as in fig. S3. The size of all the images is 1.2 nm× 1.2 nm.

As the tip height decreases, the interaction between the tip and the tetramer induces significant lateral tip relaxation for all tip models (fig. S7). The neutral and $d_{z^2}$ tips only deflect just over the dangling OH due to the Pauli repulsion force (fig. S7A and C), while the s-like tip apex surfs on the isosurface of the Pauli potential and meanwhile is strongly modulated by the attraction/repulsion of electrostatic force, giving rise to the chiral features resembling the electrostatic potential distribution (fig. S7B). In contrast, at the large tip height where the chiral electrostatic potential of tetramer is resolved with the $d_{z^2}$ tip, the lateral relaxation of the tip apex is negligible (fig. S7C, z=6.8 Å).

**VIII. Calculated electrostatic potential map of the water dimers and trimers**

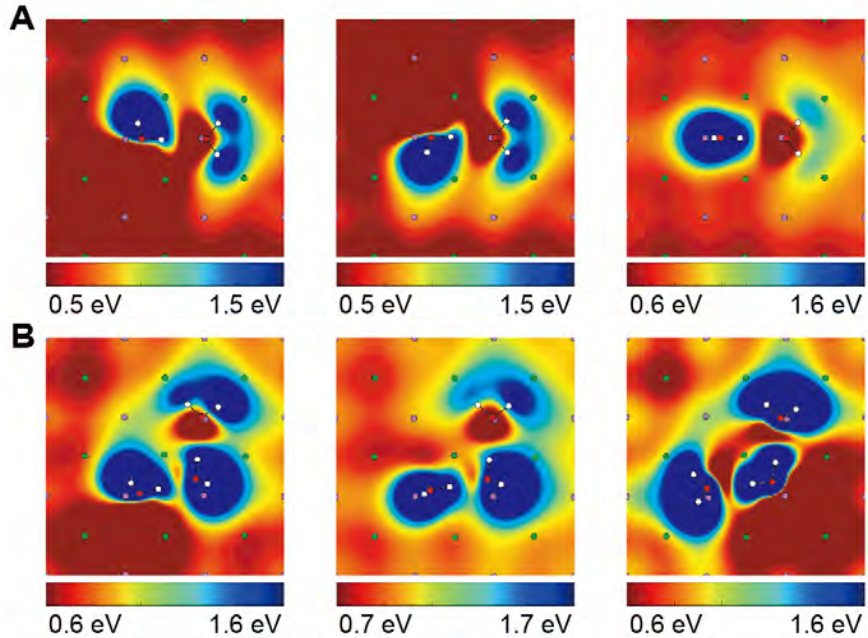

**fig. S8**. **Calculated electrostatic potential map of the water dimers and trimers.** (A)

Three water dimers corresponding to the ones in fig. 4A. (**B**) Three water trimers corresponding to the ones in fig. 4D. The plane height (which is defined as the distance from the outermost H atom) of the maps: (A) 45 pm, 45 pm, 49 pm (from left to right); (B) 23 pm, 30 pm, 14 pm (from left to right). H, O, Cl and Na atoms are denoted as white, red, green and purple dots, respectively. The size of all the images is 1.2 nm× 1.2 nm. The blue regions in the electrostatic potential maps arise from the positively charged H. From the characteristic shape and contrast of the blue features, we can easily determine the orientation of water molecules.

**IX. Energy barrier for the conversion between water dimers**

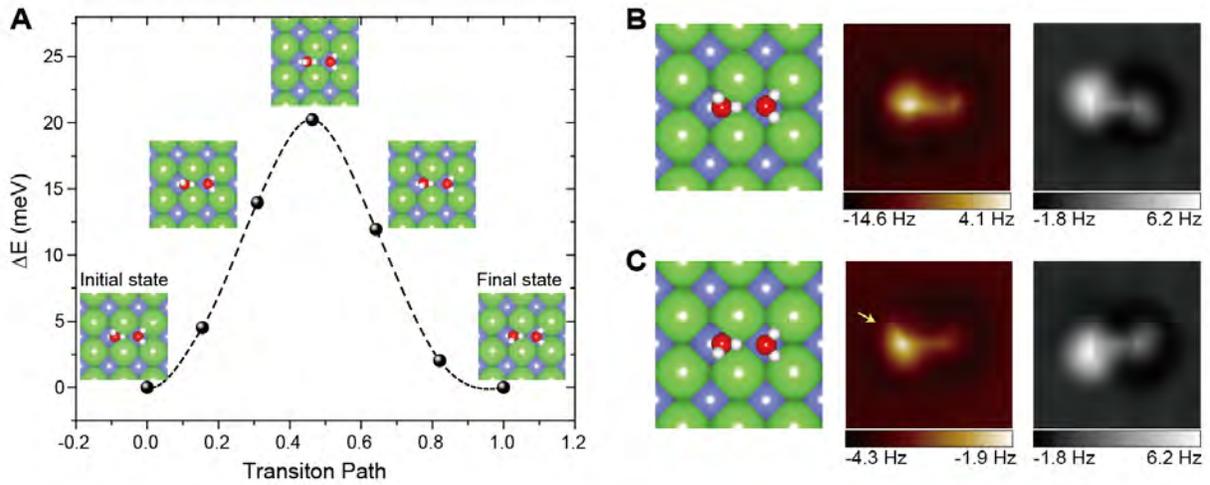

**fig. S9. Calculated energy barrier for the conversion between two degenerate water dimers.** (**A**) Transition barrier between two degenerate water dimers, which differ in the OH tilting of the left water molecule. Inset, snapshots of water dimers along the transition path. (**B** and **C**) Geometric structures, experimental and simulated AFM images of the two water dimers. The tip height of experimental AFM images is 10 pm. All the oscillation amplitudes of experimental and simulated images are 100 pm. All the simulations were done with a quadrupole ($d_{z^2}$) tip (k= 0.5 N/m, Q= -0.2e) as the water dimers were fixed. The experimental AFM images were acquired with a CO-tip. The size of the images: 1.2 nm× 1.2 nm.

As shown in fig. S9A, the transition barrier between two degenerate water dimers is so small (~20 meV) that the water dimers are extremely susceptible to the any weak perturbation induced by tip-water forces. In Fig. 4A-C, we show that the electrostatic potential of those water dimers can be stably imaged at large tip heights without inducing any disturbance, indicating the nearly non-invasive character of the AFM imaging in this region. Occasionally, we can even image the transition state of the water dimer (Fig. 4A, right). This possibility may result from the inhomogeneity of the underlying reconstructed Au substrate, which makes the transition state somewhat metastable with a marginal energy barrier. However, when scanning the water dimers at small tip heights where the short-range Pauli repulsion becomes dominant and significant tip relaxation is present, the fragile water dimers can be easily disturbed or switched (fig. S9B and C). The simulation in fig. S9B agrees well with the experimental AFM image, suggesting that the perturbation of the tip to the water dimer is negligible. In contrast, for the other type of water dimer (fig. S9C), the directionality of the upward H in the simulated image obviously deviates from the experiment result (see the yellow arrow). This discrepancy may arise from the flip of the left water molecule in the dimer under the close imaging.

## X. High-resolution AFM images of a water triple-tetramer

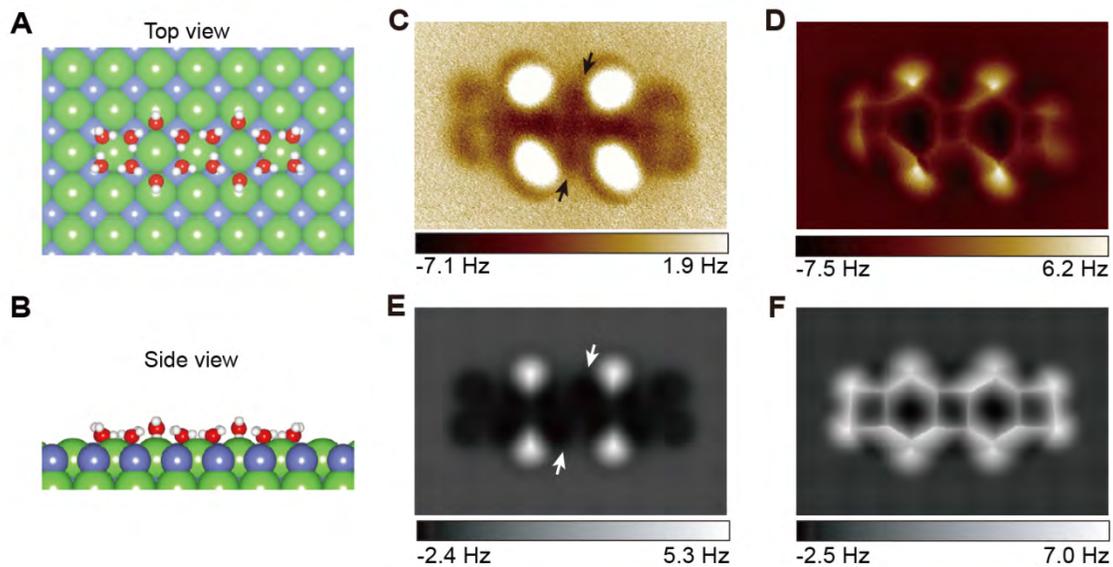

**fig. S10. High-resolution AFM images of a water triple-tetramer with a CO-tip.** (**A** and **B**) Top and side views of the atomic structures of the water triple-tetramer. (**C,D**) and (**E,F**)

Experimental and simulated Δf images of the water triple-tetramer, respectively. The tip heights: (C) 110 pm, (D) 20 pm, (E) 6.19 Å, (F) 5.16 Å. The tip height in simulations is defined as the distance between the tip apex and the highest H atom of water double tetramer. The Δf scale in (C) is adjusted to highlight the chirality of the central tetramer. The chirality of the central tetramer is denoted by arrows in (C) and (E). All the oscillation amplitudes of experimental and simulation images are 100 pm. The simulations were done with a quadrupole ($d_{z^2}$) tip (k= 0.5 N/m, Q= -0.2e). The size of the images: 2 nm× 3 nm.

The triple-tetramer is composed of three tetramers which are bridged with four standing water molecules, forming a bilayer ice cluster (fig. S10A and B) (*14*). The bridging water molecules were imaged as four bright spots at the large tip height (fig. S10C), which result from the Pauli repulsion force between the CO-tip and the standing water molecules. The chirality of the tetramers at two ends within the bottom layer of the bilayer ice can be clearly resolved, which has been not possible with STM before(*14*). Surprisingly, the chirality of the middle tetramer can be also distinguished although it is somewhat blocked by the higher bridging water molecules. The skeleton of the H-bonding network in the triple-tetramer can be seen very clearly under close imaging condition (fig. S10D). The sharp lines emerge from the deflection of the probe particle due to its repulsive interaction with the nearest neighboring water molecules (see discussions in fig. S1). Similar results were obtained for water overlayers on Cu surfaces recently(*21*). The simulated AFM images at large (fig. S10E) and small (fig. S10F) tip heights agree well with the experimental results. Note that the imaging of H-bonding skeleton requires relatively strong tip-water interaction at short range, which may induce significant disturbance to the water structure.

**XI. The disturbance of Cl-tip on the water dimer, trimer and double-tetramer**

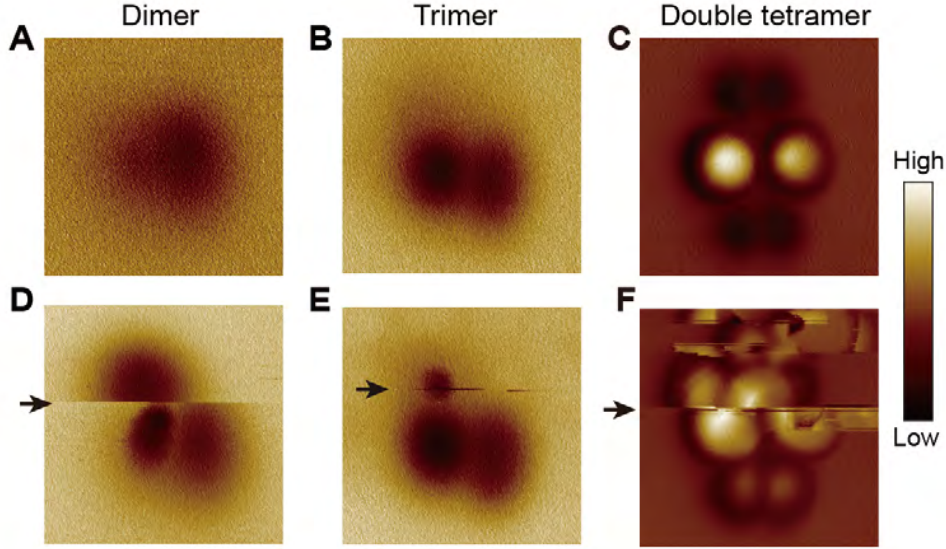

**fig. S11. The disturbance of Cl-tip on the water dimer, trimer and double-tetramer.** (**A** and **D**) Δf images of a water dimer at tip heights of 140 pm and 120 pm, respectively. (**B** and **E**) Δf images of a water trimer at tip heights of 0 pm and -30pm, respectively. (**C** and **F**) Δf images of a water double-tetramer at tip heights of -50pm and -100pm, respectively. Similar to the structure of triple tetramer (fig. S10), the double tetramer consists of two tetramers linked by two bridged water molecules. The tip height is referenced to the STM set point on the NaCl surface (100 mV, 50 pA). The oscillation amplitudes: (A), (B), (D) and (E) 100pm; (C) and (F) 50pm. The size of the images: (A), (B), (D) and (E) 1.4 nm× 1.4 nm; (C) and (F) 2 nm× 2 nm.

**Table S1 The fitted decay length of force curves obtained with different tips**

| Calculated decay length (Å) | | | Experimental decay length (Å) | |
|---|---|---|---|---|
| s tip | $p_z$ tip | $d_{z^2}$ tip | Cl-tip | CO-tip |
| 1.039±0.006 | 0.814±0.005 | 0.654±0.004 | 0.786±0.014 | 0.326±0.03 |

**Table S2 Parameters of Lennard Jones pairwise potentials for all elements**

| Element | ε [meV] | r [Å] |
|---|---|---|
| H | 0.680 | 1.487 |
| O | 9.106 | 1.661 |
| Cl | 11.491 | 1.948 |
| Na | 10.0 | 1.4 |
| Apex | 1000 | 2.000 |